\documentclass[copyright]{eptcs}
\providecommand{\event}{T. Neary, D. Woods, A.K. Seda and N. Murphy (Eds.):\\ The Complexity of Simple Programs 2008.}
\providecommand{\volume}{1}
\providecommand{\anno}{2009}
\providecommand{\firstpage}{164}
\usepackage{breakurl}

\usepackage {amsmath,amssymb,latexsym}
\usepackage {delarray,graphics,graphpap,alltt,graphicx}
\usepackage{setspace}

\newtheorem{defn}{Definition}

\newcommand{\perf}{\mathrm{Perf}}
\newcommand{\first}{\mathrm{First}}

\newcommand{\bool}[1]{\{ 0,1\}^{#1}}
\newcommand{\harp}{\upharpoonright}				

\newcommand{\N}{\mathbb{N}}

\newcommand{\np}{\ensuremath{\mathsf{NP}}}

\title{A General Notion of Useful Information}
\author{Philippe Moser
    \institute{Department of Computer Science, National University of Ireland Maynooth, Co. Kildare, Ireland.}
    \email{pmoser(at)cs.nuim.ie }
}

\def\titlerunning{A General Notion of Useful Information}
\def\authorrunning{P. Moser}

\begin{document}
\maketitle

\begin{abstract}
	In this paper we introduce a general framework for defining the depth of a sequence with respect to a class of observers.
	We show that our general framework captures all depth notions introduced in complexity theory so far.
	We review most such notions, show how they are particular cases of our general depth
	framework, and review some classical results about the different depth notions.
\end{abstract}

\section{Introduction}

	A traditional way (in the sense of algorithmic information theory) to define the complexity of a system is to look at
	the number of bits required to describe the system (its Kolmogorov complexity) i.e., the more bits needed to describe the 
	system the higher its complexity. An alternative view was proposed by Bennett in \cite{b:bennett88}, with the concept of useful information
	or logical depth. Bennett's original idea is that  any system can be either simple, random or deep; 
	where
	simple systems being completely predictable contain no useful information; random ones, being completely unpredictable, do not
	contain any useful information either; both (trivial and random) being therefore shallow objects. 
	On the other hand, systems that are neither random nor trivial i.e., that are neither fully predictable nor completely unpredictable, 
	contain useful information; they are called deep objects. 
	Although random sequences contain a 
	lot of information (in the sense of algorithmic information theory), this information is not
	of much value, and such sequences are shallow. Contrast this with a ten days weather forecast; although it contains no more information than 
	the differential equations from which it was originally simulated (i.e. the number of bits to describe the system is small), it saves its owner 
	the effort of running the (costly) simulation again.

	Bennett argued that deep structures, because they contain complex well-hidden patterns, cannot be created by short computations.
	This observation was formalized in the so-called \emph{slow growth law}, which states that no  short computation (truth table reduction)
	can transform a shallow sequence into a deep sequence.

	Bennett's logical depth is based on Kolmogorov complexity. Intuitively, a binary sequence is deep, if the 
	more time an algorithm is given, the better it can compress the sequence. Although Bennett's formulation
	is theoretically very elegant, it is uncomputable, due the uncomputability of Kolmogorov complexity.

	To overcome the uncomputability of logical depth, several notions of feasible depth have been proposed so far.
	In \cite{DBLP:journals/iandc/LathropL99} Juedes, Lathrop and Lutz proposed a recursive version of depth, called recursive computational depth.
	Although recursive depth gets rid of  the uncomputability in Bennett's formulation,
	it is still far from feasible and cannot be used
	in complexity theory, as most languages of interest in complexity theory --e.g. any $\np$-complete language-- being computable,
	are neither recursive-deep nor Bennett deep. 
	This strongly motivates the study of polynomial depth notions in complexity theory, and several such notions have been	
	studied so far. Antunes, Fortnow, Melkebeek and Vinodchandran introduced a depth notion based on polynomial
	distinguishing complexity in  \cite{b.antunes.depth.journal}. Doty and Moser studied a depth notion based on polynomial time
	predictors \cite{DBLP:conf/cie/DotyM07}, and a notion based on polynomial monotone compressors was introduced in \cite{b.moser.monotone.depth}. 
	In an attempt to reduce the time bounds
	even further, Doty and Moser \cite{DBLP:conf/cie/DotyM07} introduced a finite state versions of depth (based on finite-state compressors).

	As seen by the many notions of depth that were studied, depth is not an absolute but a relative notion i.e.,
	a deep sequence is always deep \emph{with respect} to some class of observers. As an example a book of Chinese poetry
	will look like a sequence of randomly ordered symbols to an observer that does not read Chinese i.e., not deep from his point of view. 
	On the other hand,
	for Chinese readers, the more they read and study the book, the more information they will be able to extract from it, therefore considering it 
	deep from their point of view.

	In this paper we introduce a general framework for defining the depth of a sequence with respect to a class of observers $G$.
	Informally we say a sequence is deep relative to $G$ (or $G$-deep) if for any observer $A$ from $G$ there is an observer $A'$ from $G$ such that
	$A'$ can extract more information from $S$ than $A$ can. 
	More formally suppose we have a class $G$ of algorithms that compute on binary strings, together with a function
	$\perf$ that measures (by a number between 0 and 1) how well an algorithm $A$ in $G$ performs on an input string $x$. For example,
	if $G$ is a class of compression algorithms, then given a compression algorithm $A$ in $G$ and a string 
	$x$, a possible performance measure of $A$ on $x$ is the compression ratio $\frac{|A(x)|}{|x|}$. As another example
	suppose $G$ is a class of predictors, trying to predict the bits of the characteristic sequence of an $\np$-complete
	language like SAT, a possible performance measure is the number of correct predictions divided by the total number of bits.
	The class $G$ together with the performance function yields a natural notion of depth relative to $G$ i.e., a definition
	of useful information in the eye of observers drawn from class $G$.

	Similarly to Bennett's notion, both simple and random (with respect to $G$)
	sequences are not deep relative to $G$; indeed if a sequence is simple with respect to $G$ meaning there is an observer in $G$ that can
	fully describe the sequence, then no observer in $G$ can do any better i.e., the sequence is not $G$-deep. 
	In the same vein, if a sequence is random with respect to $G$,
	then no observer in $G$ can predict the sequence, hence all observers in $G$ are equally bad at predicting it i.e., the sequence is not 
	$G$-deep. 

	As we show in this paper, our general framework actually captures all depth notions introduced in complexity theory so far 
	\cite{b:bennett88,DBLP:journals/iandc/LathropL99,b.antunes.depth.journal,DBLP:conf/cie/DotyM07,b.moser.monotone.depth}, which can all be seen
	as particular instances of our general depth framework. Most of these notions, are based on some class of compression algorithm,
	which as seen by our general depth framework, is only one --among many others-- particular way to define the depth of a sequence,
	therefore leaving the door open to the study of new depth notions, not necessarily based on the  compression paradigm.

	The rest of the paper is organized as follows: in Section \ref{s.gen.def} we introduce our general depth framework. In Section \ref{s.review.depth.notions},
	we review most depth notion that were introduced so far in complexity theory, show how they are particular cases of our general depth
	framework, and review some classical results about the different depth notions.

\section{Preliminaries}	 
	We write $\mathbb{N}$ for
	the set of all nonnegative integers.
	Let us fix some notations for  strings and languages.
	A \emph{string} is an element of $\bool{n}$ for some integer $n$.
	For a string $x$, its length is denoted by $|x|$. The empty string is
	denoted by $\lambda $. We denote by $\lambda,s_0,s_1,s_2,\ldots$ the standard
	enumeration of strings in lexicographical order. For a set of strings $S$,
	let $\first(S)$ denote the first (with respect to the lexicographical order)
	string of $S$. We say string $y$ is a prefix of string $x$, denoted $y\sqsubset x$ (also $y\sqsubseteq x$), 
	if there exists a string $a$ such that
	$x=ya$. A set of strings is prefix free if no  string in the  set is the prefix of another string in the set.

	A sequence is an infinite binary string, i.e. an element of $\bool{\infty}$.
	For $S$ $\in $ $\bool{\infty} $ and $i,j$
	$\in $ $\mathbb{N}$, we write $S[i.. j]$ for the string consisting of
	the $i^{\textrm{th}}$ through $j^{\textrm{th}}$ bits of $S$, with
	the convention that $S[i.. j]=\lambda $ if $i>j$, and $S[0]$ is the
	leftmost bit of $S$. We write $S[i]$ for $S[i.. i]$ (the
	$i^{\textrm{th}}$ bit of $S$). Unless otherwise specified,
	logarithms are taken in base $2$.

	A \emph{language} is a set of strings.
	The characteristic sequence of a language $L$ is the sequence $\chi_L \in\bool{\infty}$,
	whose $n$th bit is one iff $s_n \in L$. We will often use the notation $L$ for $\chi_L$.

	TM stands for Turing machine. A monotone TM is a TM such that for any strings $x,y$, $M(x)\sqsubseteq M(xy)$.
	A TM is called prefix free if the set of admissible programs is a prefix free set.

\section{A general notion of depth}\label{s.gen.def}

	Before giving the formal definition of our depth framework, let us generalize the informal description given in the 
	introduction to two classes of observers instead of one i.e., 
	suppose we have two classes $G,G'$ of algorithms that compute on binary strings, together with a function
	$\perf$ that measures (by a number between 0 and 1) how well an algorithm $A$ in $G\cup G'$ performs on an input string $x$.

	The classes $G,G'$ together with the performance function yields a natural notion of depth relative to $G,G'$, i.e. a definition
	of useful information in the eye of observers competing against each other and drawn from class $G$ and $G'$. 
	Informally a sequence $S$ is $(G,G')$-deep, if for any
	algorithm $A$ in $G$ there is an algorithm $A'$ in $G'$ such that $A'$ performs better on $S$ than $A$.
	
	More formally let $G,G'$ be two classes of algorithms computing on binary strings. Let 	
	$\perf: (G\cup G')\times\bool{*}\rightarrow [0,1]$ be a function that measures how well an algorithm $A$ in 
	$G\cup G'$ performs on an input string $x$, where $1$ (resp. 0) means optimal performance  (resp. worst performance).
	Usually $G'$ is at least as powerful (or equal to) as $G$ (formally for every $A\in G$ there exists $A\in G'$ such that for any string $x$, 
	$\perf(A',x)\geq \perf(A,x)$).
	Let $M$ be a family of polynomial time computable functions, where for every $m\in M$ and
	every integer $n$, $1\leq m(n) \leq n$; for example  $M= O(\log n)$. $M$ will measure by how much an algorithm $A'$ performs
	better than algorithm $A$. 
	We thus have all the tools to define depth with respect to $(G,G')$.
	\begin{defn}
		A sequence $S\in\bool{\infty}$ is $(G,G')$-deep if there exists a bound $m\in M$ such that for every $A\in G$ there exists $A'\in G'$ 
		such that for infinitely many $n\in\N$
		$$\perf(A',S[1..n]) - \perf(A,S[1..n]) \geq \frac{m(n)}{n}.$$
	\end{defn}
	Other variations that have been considered are obtained by replacing ``there exists a bound $m\in M$'' by ``for all bounds $m\in M$'',
	and/or  ``for infinitely many $n\in\N$'' by ``for almost every $n\in\N$'' (the latter being sometimes referred to as strongly deep).
	However subtle the influence of those small changes on the corresponding depth notion is,  they are not relevant to the high level understanding 
	of the notion of depth we aim at in the present paper, and can all be seen as variations of a same common
	theme, which is captured by our general depth framework, and that  in essence says that 
	a sequence is deep with respect to $(G,G')$ if given any algorithm in $G$ there is an algorithm in $G'$
	that performs better (as measured by $M$) on the sequence. 

	This general definition actually captures all depth notions introduced in complexity theory so far 
	\cite{b:bennett88,DBLP:journals/iandc/LathropL99,b.antunes.depth.journal,DBLP:conf/cie/DotyM07,b.moser.monotone.depth}, 
	as we shall see in the next section,
	where
	we will review some of these notions together with some of the results that were obtained for each of them.

\section{A review of some computational depth notion}\label{s.review.depth.notions}

\subsection{Bennett's logical depth}

	The first notion of depth, Bennett's logical depth 
	\cite{b:bennett88}, 
	is based on Kolmogorov complexity.
	We rephrase it in our general depth framework. We need the following broad definition of 
	compressor.
	\begin{defn}
		A compressor is a one-to-one function $C:\bool{*}\rightarrow\bool{*}$.
 	\end{defn}
	A compressor is computable if there is a TM $T$ such that for any string $x$, $T(x)=C(x)$.
	For the rest of this paper fix a prefix-free universal TM $U$. The universal compressor is given by
	$$C_U(x)= \first \{p| \ U(p)=x \}$$
	i.e. its the first (hence shortest) program that makes $U$ output $x$.
	It is well known that the choice of $U$ affects the definition of $C_U$ only up to an additive constant. We therefore 
	omit $U$ in the notation and write $C_K$.
	Bennett's logical depth is obtained by letting
	\begin{equation*}
	\begin{split}
	G & = \{C|\ C \text{ is a computable compressor}\}\\
	G' & = \{C_K\}\\
	\perf(C,x) & =1-\frac{|C(x)|}{|x|} \\
	M & =\N 
	\end{split}
	\end{equation*}
	i.e. $S$ is Bennett-deep if 
	for every $C\in G$, every $m\in\N$ and almost every $n\in\N$
	$$|C(S[0..n])| - |C_K(S[0..n])| \geq m.$$

	Among others, Bennett showed in \cite{b:bennett88} that both Martin-L\"of random and recursive sequences are shallow (i.e. not deep), 
	logical depth satisfies a slow growth law for truth-table reductions, and that the Halting language is deep.
	This was later generalized by Juedes, Lathrop  and Lutz \cite{DBLP:journals/tcs/JuedesLL94} 
	to the class of weakly useful languages --a language is weakly useful if the set of languages reducible to it is not small
	(in a computable Lebesgue measure sense; see \cite{DBLP:journals/tcs/JuedesLL94} for more details)-- 
	where it was shown that every weakly useful language is Bennett-deep.

	The main limitation of Bennett's notion is that the universal compressor $C_K$
	is not computable. A way to overcome this was proposed by Juedes, Lathrop and Lutz in \cite{DBLP:journals/iandc/LathropL99}, by replacing $G'$ with $G$,
	in Bennett's formulation, which we describe in the next section.	

\subsection{Recursive computational depth}

	Recursive computational depth was proposed in \cite{DBLP:journals/iandc/LathropL99} as a way to overcome the uncomputability in Bennett's definition.
	It is obtained by letting 	\begin{equation*}
	\begin{split}
	G & = G' = \{C|\ C \text{ is a computable compressor}\}\\
	\perf(C,x) & =1-\frac{|C(x)|}{|x|} \\
	M & =\N 
	\end{split}
	\end{equation*}
	i.e. $S$ is recursive deep if 
	for every $C\in G$, and every $m\in\N$ there exists $C'\in G$ such that for almost every $n\in\N$
	$$|C(S[0..n])| - |C_K(S[0..n])| \geq m.$$

	Among others it was shown in \cite{DBLP:journals/iandc/LathropL99} that Bennett's depth and recursive depth are two separate notions,  that recursive depth also satisfies 
	a slow growth law for truth-table reductions and that both Martin-L\"of random and recursive sequences are not recursive deep.

	Although recursive depth gets rid of  the uncomputable universal compressor $C_K$, it is still far from feasible and cannot be used
	in complexity theory, as most languages of interest in complexity theory --e.g. any $\np$-complete language-- being computable,
	are neither recursive-deep nor Bennett deep. This motivates a notion of polynomial depth in the context of complexity theory.
	We review the first such notion in the following section.

\subsection{Distinguishers based polynomial time depth}

	Antunes, Fortnow, Melkebeek and Vinodchandran introduced three depth notions in \cite{b.antunes.depth.journal}. 
	The first two are only ``half polynomial'' in the sense that $G'$ contains the universal compressor
	$C_K$. The third notion was obtained via the notion of distinguisher. Here is a definition.
	\begin{defn}
		A distinguisher $D$ for a string $x$ is a  function $D:\bool{*}\rightarrow\bool{}$, such that for any string~$z$,
		$$D(z) = 
		\begin{cases} 
		1 & \text{if } z=x \\
		0 & \text{otherwise.} 
		\end{cases}
		$$
		A polynomial time distinguisher for $x$ is a distinguisher $D$ for  $x$ that is computable in time polynomial in~$x$.
 	\end{defn}
	The third depth notion in \cite{b.antunes.depth.journal} is obtained by letting 
	\begin{equation*}
	\begin{split}
	G & =  \left\{D| D(x) = \first \{p| \ U(p)=x \text{ in time polynomial in $|x|$}\}\right\} \\
	G' & =\left\{ C| C(x) = \first \{p| \ U(p,\_ ) \text{ is a polynomial distinguisher for $x$}\}\right\} \\
	\perf(A,x) & =1-\frac{|A(x)|}{|x|} \quad\text{for any $A\in G\cup G'$} \\
	M & = O(\log n).
	\end{split}
	\end{equation*}
	The relationship between $G$ and $G'$ i.e. between distinguishing a string and producing a string in polynomial time is not known,
	(beyond the fact that producing implies distinguishing).

	Connections were demonstrated between depth and average-case complexity, nonuniform circuit complexity, and efficient search 
	for satisfying assignments to Boolean formulas in \cite{b.antunes.depth.journal}, and the third depth notion was shown to satisfy a slow growth law for restricted polynomial time reductions.

	As mentioned earlier, the two first notions have non-computable $G'$. For the third notion
	it is currently not known whether $G$ and $G'$ are computable in polynomial time, i.e. given $A\in G$ (or $G'$), it is not known if $A$ is
	polynomial time computable (we only know that $A$ is exponential time computable). 
	An attempt to overcome this difficulty was proposed by Doty and Moser in \cite{DBLP:conf/cie/DotyM07}, by considering polynomial 
	time predictors. We review their notion in the following section.

\subsection{Predictors based polynomial depth}

	The depth notion from Doty and Moser \cite{DBLP:conf/cie/DotyM07} is based on polynomial time oblivious predictors, 
	that try to predict the next bit of the characteristic sequence of a language
	without having access to the history of previously seen bits; here is a definition
	\begin{defn}	
		An \emph{oblivious predictor} is a function $p:\bool{*}\times \bool{} \rightarrow [0,1]$ such that, for all $x\in\bool{*}$, $p(x,0)+p(x,1)=1$. 
	\end{defn}
	Intuitively, when trying to predict a language $L$, $p(x,1)$ is the probability with which the predictor predicts that $x\in L$. To measure how well a predictor $p$ 		
	predicts $L$, we consider 			
	its associated martingale $p:\bool{*}\rightarrow [0,\infty)$ given by
    $$d_p(L\harp n)= 2^{n}\prod_{y \leq s_n} P(y,L(y)).$$
	This definition can be motivated by the following  betting game in which  gambler $d_p$
	puts bets on the successive membership bits of the hidden language $L$.
	The game proceeds in infinitely many rounds where at the end of round $n$,
	it is revealed to the gambler whether $s_{n} \in L$ or not.
	The game starts with capital 1.
	Then, in round $n$, $d_p$ bets a certain fraction
	$\epsilon_{w}p(w)$ of his current capital $d_p(w)$, that the $n$th word
	$s_{n} \in L$, and bets the remaining capital $(1-\epsilon_{w})d_p(w)$ on
	the complementary event $s_{n} \not \in L$.
	The game is fair, i.e. the amount put on the correct event is doubled,
	the one put on the wrong guess  is lost.
	The value of $d_p(w)$, where $w = \chi_{L}[0.. n]$ equals the
	capital of $d_p$ after round $n$ on language $L$.

	A polynomial predictor $p$ is a predictor such that $p(s_n,b)$ is computable in time polynomial in $n$. Polynomial predictors were used in
	\cite{DBLP:conf/cie/DotyM07} to define a polynomial depth notion by letting 
	\begin{equation*}
	\begin{split}
	G & = G'=\{p| \text{ $p$ is a polynomial time predictor}\}\\
	\perf(p,x)&=\log d_p \\
	M&=\{\log\log n +O(1)\}
	\end{split}
	\end{equation*}	
	It was shown in \cite{DBLP:conf/cie/DotyM07} that this depth notion satisfies a slow growth law for restricted polynomial time reductions,  
	that similarly to Bennett's notion,
	polynomial time weakly useful languages (a polynomial version of weakly useful languages) are polynomial deep, and that the corresponding 
	polynomial time version of random and computable sets are not polynomial deep.

	Polynomial oblivious predictors are somehow restricted because they cannot access the  history of previously seen bits. As an attempt to overcome
	this limitation, a notion of depth based on monotone polynomial time compressors was introduced in \cite{b.moser.monotone.depth}, where it was shown that this depth 
	notion satisfies a slow growth law for restricted polynomial time reductions,  that the corresponding 
	polynomial time version of random and computable sets are not polynomial deep, and that the set
	of Levin random strings is deep in the sense of monotone polynomial depth.
	
	Secondly, although the predictors are polynomial time computable, we have no control over the polynomial exponent. In an attempt to reduce the computational power
	of the algorithms even further, a finite state version of depth was introduced by Doty and Moser in \cite{b.moser.monotone.depth}, which we review next.

\subsection{Finite-state Depth}

	Finite-state depth \cite{DBLP:conf/cie/DotyM07} is based on the standard model of finite-state transducer.
	A \emph{finite-state transducer (FST)} is a 4-tuple $T = (Q,\delta,\nu,q_0),$ where
\begin{itemize}
    \item $Q$ is a nonempty, finite set of \emph{states},

    \item $\delta: Q \times \bool{} \to Q$ is the \emph{transition function},

    \item $\nu: Q \times \bool{} \to \bool{*}$ is the \emph{output function},

    \item $q_0 \in Q$ is the \emph{initial state}.
\end{itemize}
As usual we consider the canonical extension of the transition function, i.e.
for all $x\in\bool{*}$ and $a\in\bool{}$, define  $\widehat{\delta}:\bool{*} \to Q$ by the recursion
    $$\widehat{\delta}(\lambda) = q_0, \text{ and }\widehat{\delta}(xa) = \delta(\widehat{\delta}(x),a).$$

For $x\in\bool{*}$, we define the \emph{output} of $T$ on $x$ to be the string $T(x)$ defined by the recursion
    $$T(\lambda) = \lambda, \text{ and } T(xa) = T(x)\nu(\widehat{\delta}(x),a)$$
for all $x\in\bool{*}$ and $a\in\bool{}$.

An FST can trivially act as an ``optimal compressor'' by outputting $\lambda$ on every transition arrow, but this is, of course, a useless compressor, because the input cannot be recovered. An FST $T = (Q,\delta,\nu,q_0)$ is  \emph{information lossless (IL)} if the function $x \mapsto (T(x),\widehat{\delta}(x))$ is one-to-one; i.e., if the output and final state of $T$ on input $x\in\bool{*}$ uniquely identify $x$. A finite state compressor, is an \emph{information lossless finite-state transducer} (ILFST). We write ILFST to denote the set of all information lossless finite-state transducers.

	It is a remarkable fact \cite{Huff59a,Koha78} that  the function from $\bool{*}$ to $\bool{*}$ computed by an ILFST can be inverted -- in an approximate sense -- by
	another ILFST, i.e. if the compression can be done by an ILFST then there exists a decompressor that is also an ILFST.

	The finite state depth notion from \cite{DBLP:conf/cie/DotyM07} is obtained by considering finite state compressors, i.e. by letting 
	\begin{equation*}
	\begin{split}
	G & = G'= \mathrm{ILFST}\\
	\perf(C,x) & = 1-\frac{|C(x)|}{|x|}\\
	M & = \{m(n)=\alpha n| \ n\in[0,1]\}
	\end{split}
	\end{equation*}

	It is shown in \cite{DBLP:conf/cie/DotyM07} that FS-deep sequences exist, that FS-depth satisfies a slow growth law for information lossless finite state transducers,
	and that the corresponding 
	finite-state versions of random and computable sets are not finite-state deep.

\section{Conclusion}
	We proposed a general framework for defining the depth of a sequence, that captures all existing notions of depth
	studied so far in complexity theory. Although most depth notions studied so far concentrated on the compression point of
	view, it is only one particular case of the more general depth framework proposed here. We therefore anticipate the arising of new
	depth notions --not necessarily compression algorithm based-- that will shed new light on the theory of computational depth.

\bibliographystyle{eptcs}

\begin{thebibliography}{}
\providecommand{\bibitemstart}[1]{\bibitem{#1}}
\providecommand{\bibitemend}{}
\providecommand{\bibliographystart}{}
\providecommand{\bibliographyend}{}
\providecommand{\url}[1]{\texttt{#1}}
\providecommand{\urlprefix}{Available at }
\providecommand{\bibinfo}[2]{#2}
\bibliographystart

\bibliographyend
\end{thebibliography}


\begin{thebibliography}{1}

\bibitem{b.antunes.depth.journal}
L.~Antunes, L.~Fortnow, D.~van Melkebeek, and N.~Vinodchandran.
\newblock Computational depth: Concept and applications.
\newblock {\em Theoretical Computer Science}, 354:391--404, 2006.

\bibitem{b:bennett88}
C.~H. Bennett.
\newblock Logical depth and physical complexity.
\newblock {\em The Universal Turing Machine, A Half-Century Survey}, pages
  227--257, 1988.

\bibitem{DBLP:conf/cie/DotyM07}
D.~Doty and P.~Moser.
\newblock Feasible depth.
\newblock In S.~B. Cooper, B.~L{\"o}we, and A.~Sorbi, editors, {\em CiE},
  volume 4497 of {\em Lecture Notes in Computer Science}, pages 228--237.
  Springer, 2007.

\bibitem{Huff59a}
D.~A. Huﬀman.
\newblock Canonical forms for information-lossless ﬁnite-state logical
  machines.
\newblock {\em IRE Trans. Circuit Theory CT-6 (Special Supplement)}, pages
  41--59, 1959.

\bibitem{DBLP:journals/tcs/JuedesLL94}
D.~W. Juedes, J.~I. Lathrop, and J.~H. Lutz.
\newblock Computational depth and reducibility.
\newblock {\em Theor. Comput. Sci.}, 132(2):37--70, 1994.

\bibitem{Koha78}
Z.~Kohavi.
\newblock Switching and finite automata theory (second edition).
\newblock {\em McGraw-Hill}, 1978.

\bibitem{DBLP:journals/iandc/LathropL99}
J.~I. Lathrop and J.~H. Lutz.
\newblock Recursive computational depth.
\newblock {\em Inf. Comput.}, 153(1):139--172, 1999.

\bibitem{b.moser.monotone.depth}
P.~Moser.
\newblock The set of {L}evin {K}t-random strings is polynomially deep.
\newblock {\em submitted}, 2008.

\end{thebibliography}

\end{document}